# Responsibility and Regulation: Exploring Social Measures of Trust in Medical AI

## Glenn McGarry

School of Computer Science, University of Nottingham, Nottingham, United Kingdom, glenn.mcgarry@nottingham.ac.uk

## Andy Crabtree

School of Computer Science, University of Nottingham, Nottingham, United Kingdom, andy.crabtree@nottingham.ac.uk

## Lachlan Urquhart

School of Law, University of Edinburgh, Edinburgh, United Kingdom, lachlan.urquhart@ed.ac.uk

## Alan Chamberlain

School of Computer Science, University of Nottingham, Nottingham, United Kingdom, alan.chamberlain@nottinghm.ac.uk

This paper explores expert accounts of autonomous systems (AS) development in the medical device domain (MD) involving applications of artificial intelligence (AI), machine learning (ML), and other algorithmic and mathematical modelling techniques. We frame our observations with respect to notions of responsible innovation (RI) and the emerging problem of how to do RI in practice. In contribution to the ongoing discourse surrounding trustworthy autonomous system (TAS) [29], we illuminate practical challenges inherent in deploying novel AS within existing governance structures, including domain specific regulations and policies, and rigorous testing and development processes, and discuss the implications of these for the distribution of responsibility in novel AI deployment.

## 1 INTRODUCTION

As the integration of intelligent technologies in everyday life becomes an ever more realistic prospect, so too are concerns surrounding balancing the benefits of these innovations to individuals and society against a host of ethical, legal, social, and economic concerns [32]. Cross-organizational working groups and academic discourse have sought to conceive frameworks designed to underpin trust in automated systems [5, 27] and iteratively develop values-based approaches to research and innovation (R&I) governance that is "concerned with aligning societal values on the one hand with developments in science and technology on the other" [2]. For example, originating from the political sphere of the EU the

ELSI agenda brought ethical, legal, and societal impact of R&I into focus, from which the Responsible Research and Innovation program (RRI) developed [19] that problematises current innovation practices and establishes dimensions for aligning R&I's products and processes with societal needs, values, and expectations [20, 26]. Establishing these agendas, however, has been far from straightforward with much of the academic world "scratching their heads" [12] on how to define its dimensions unambiguously. Further still, with little practical experience in this space amongst the AI community, new problems have opened for the development of next generations of intelligent technologies, as we discuss next.

## 1.1 How to 'Do' RI

The requirement for researchers to adhere to RI's principles are increasingly being written-in to the mandatory criteria of publicly funded research programs through, for example the AREA framework [4], which bids researchers to Anticipate, Reflect, Engage, and Act in terms of the societal, ethical, and environmental impacts of their R&I activities. This obligation is driving a second wave of research that attempts to interpret, study, and understand just how to *do* RI [13], insights which are largely absent from principled frameworks related to AS [1, 9, 28]. Schuijff et al [21] attempt to bridge this gap through studies of RI in practice while calling for improved descriptions of RI methodologies [21]. Attempts to support researchers include the development of RRI cards that provide prompts for considering different dimensions of the AREA framework, however as their creators point out it "cannot be considered that RI is done" by engaging with them [14]. Further to this Portillo et al's [15] studies of researcher engagement with RRI tools, such tools as the ORBIT RRI self-assessment and moral-IT cards [2, 21, cited in 12], concludes that institutional commitment to RI and support for researchers is crucial, particularly when faced with new generations of legislation such as the EU's proposed AI specific legislation [3].

## 1.2 The Governance Dilemma

RRI largely developed in response to established innovation practices, in which successes or failures are determined in the marketplace and often subject to governance only *after* the risks to society and the environment have emerged. Further still, risk-based governance often does not account for the impact of novel technologies beyond the concern for producing safe products, effectively rendering innovation as a potentially harmful enterprise [26]. This gives rise to a governance dilemma: how to minimize the prospective harms of innovation without suppressing the potential benefits to humanity? RRI addresses this with a call for new modes of innovation governance that incorporate ethical and social concerns including anticipatory governance in the form of technology and impact assessment [17, 24]; a shift from risk governance to innovation governance; the adoption of standards/codes of conduct (and the development of new ones) to ensure market accountability; governance structures for "incorporating ethical principles in the design process"; and models of governance that stipulate mechanisms for co-responsibility of actors [20]. These concerns are shared at the level of international executive bodies including the UN [29], and the EU [3], while research communities call for greater accountability for the products of AI [18, 30].

## 2 THE STUDY

To explore these concerns, we engaged with participants recruited for their expertise in the design and deployment of novel medical devices (MD) intended to provide clinician support and improve healthcare economics within a range of clinical settings. Such devices increasingly use AI in shifting from relatively simple categorisation problems (e.g., this image is type A, that type B) to diagnosis of medical conditions and must be approved by a regulator in the various countries in which they are sold. This includes explicit consideration of AI-based features and the current regulatory environment is described as "very risk aversive", particularly in a context where the introduction of machine learning changes the "intended purpose" of devices, taking on (but not necessarily supplanting) human jobs of work, and where there are no prerequisite devices against which to benchmark accuracy and clinical efficacy [10].

## 2.1 Practicalities

Our study consists of a series of four, 60-minute interviews using the telepresence application MS Teams. Each interview was an open-ended discussion designed to encompass the topics of designing for margins of error; quality management and development processes. Two interviews feature P1, a software specialist for a MD regulator investigating the regulation of software as a medical device (SaMD) - a subset of MDs that stand-alone from a specific hardware [11]; and two interviews featured three members of a research and development team for an international MD vendor: P2, The R&D Team Leader; P3 Data Scientist; and P4 Project Manager, all of whom had been involved in successfully developing, integrating, and marketing a novel ML-driven features to support clinicians with simple image classification. Some parts of the discussions were of a commercially sensitive nature and as such the data gathered are subject to a non-disclosure agreement (NDA) that precludes certain details of conversations reported here, for example, commercial brands and specific functionalities.

## 2.2 The Findings

To begin we give a brief insight into the organisation of medical device regulation (MDR) and its processes. Regulatory authorities are "gatekeepers" to the markets in the different geographical areas that they administer - for example, the MHRA in the UK, or the FDA in the US – and are responsible for prosecuting compliance failures; interpreting legal frameworks and standards and issuing guidance for manufacturers; and handling the reporting of adverse events. In premarket systems (US and Canada), all certification is done in-house under tight regulatory control. Post-market systems (UK and EU) devolve responsibility for premarket compliance assessments and certification (EU 'CE', or UK 'CA' mark) – to approved bodies leaving the regulator to focus resources on post-market monitoring [33].

This difference of approach has significant implications for Medical AI. For premarket authorities, competition for the hire of AI expertise is problematic, while post-market systems allow some flexibility to outsource the regulatory assessment of novel products. However, as P1 explains, regardless of the regulatory model Medical AI still falls between the cracks due to the rate of in-service software changes, such as security patches, usability updates, or improved ML data sets being incompatible with the speed of regulatory change approval processes. As P1 puts it: *"getting this to run smoothly at a pace that the software development and the AI communities are comfortable with - there's a gap there … (that is) not ideal for continuous learning products"*. The challenges for Medical AI, however, begin much earlier in the process.

### 2.2.1 Entering the MD market

Entry to the MD market is not exclusive to established manufacturers, such as our participants, and may include clinicians with a proposed solution, technologists with products applicable to health care, or 'spinout' enterprises from universities. For newcomers, however, the barriers to entry are high and dependent on a specific set of understandings: *"…you need to understand what good design engineering and manufacturing practices are… and you need to understand how healthcare works, … and then (understanding medical device) regulation."* (P1). It is not adequate for technical experts to, *"run around trying to find nails for their hammer, rather than designing the right tool"* (P1), without aligning understandings of a clinical problem to prospective technological capabilities. Developers must first apply to the regulator with a well-defined intention for use statement, which is the crucial starting point for a product's journey through the regulatory process to market. As P1 puts it *"you could say 'this is for monitoring all aspects of diabetes' … but that'll take you 20 years to* *get enough evidence to satisfy that there are no gaps in safety"* – thus the more well-defined the product's intended use is, the more feasible it is to evidence its safety and effectiveness. Innovators should also take steps to understand the professional community's desire for innovative interventions and what factors might shape its design, as P2 explains: *"it's not ML that's the issue ... the ML community, has come in from the side and pushed this idea that 'hey I think our system can actually replace the doctor!', which I can tell you was not the right way to enter the market"*.

### 2.2.2 Risk is everywhere

Regulatory classification of MDs is based on inherent risk, for example tongue depressors or syringes are low risk; x-ray or laser imaging devices are high risk; and in vitro devices such as heart pacemakers are very high-risk devices - classes I, II, and III respectively in the examples above. Risk classification drives the entire process developing and marketing new MDs, as P2 explains *"Risk is everywhere. The first thing to do is (identify) the class of risk, based on the intention for use of the device. ... the biggest difference is between Class I going to Class II ... there's so much more documentation required to demonstrate that you've taken risk into account."*.

For novel AI, risk-based regulation is problematic as classifications often rely on precedent products to inform and qualify their inherent risks, and may even straddle multiple risk classifications requiring conformity assessment to the highest risk category and thus developed at higher cost. Further stil, as P1 states: *"the concept of automation isn't in the classification rules beyond broad concepts like whether you are informing or driving treatment. ... That is the human in the loop/human out of the loop/human on the loop in the AI world"*. This gives way to a dilemma in which the full extent of risk from novel technologies might be ambiguous; the classification and treatment of risk difficult to qualify; and the manifold regulatory pathways to market rendered indistinguishable. Addressing this requires honing down product scope supported by relevant scientific, technical literature, case studies; and understanding and managing risks through good design engineering and manufacturing practices that turn upon attending to quality management standards.

### 2.2.3 Rigorous processes result in rigorous documentation and rigorous products

MDR requires vendors to uphold good design, development, and manufacturing to recognised ISO standards [6–8], which includes implementing an auditable quality management system (QMS) to ensure standards of quality and safety are built into the processes and all design activities and decisions are documented and risk assessed. As P1 puts it: *"the majority of the risks in a product come from the design journey it has taken' ... The concept is that rigorous processes result in rigorous documentation and rigorous products"*. This is reinforced by P4: *"... those phases for developing something new or improving a feature in a device is coupled with our QMS, because it tells us what we need to do to show that we're taking account of the potential risks, how we're capturing them, how we intend to mitigate them, and how we intend to test those mitigations. So, it's not a hap-hazard thing"*. Adhering to these development standards is central to business practices surrounding the full lifecycle of the product, and crucial to the production of MDs that are safe by design. Compliance with quality management standards represents a heavy load on business resources and does not guarantee certification for market, particularly for novel solutions without precedent for which real-world prospective trials may be required.

### 2.2.4 The importance of prospective trials: a system is not always equal to the sum of its parts

While technical accuracy of MDs is the gold standard for safety integrity, establishing an acceptable accuracy can be problematic particularly for novel devices where *"there is no device against which (it) can be compared"* (P3). Per regulatory requirements, vendors must establish an agreed benchmark for their device's accuracy, which is complicated due to the social factors that influence the outcomes.

Firstly, errors can be costly in terms of patient harm and economic cost. As P3 points out *"There is a trade-off ... if you could adjust your system so that it gave less false negatives, the result would be more false positives, (but) ... if a patient is coming back year after year with a slow progressive disease ... then a false negative (diagnosis) is maybe not as severe as you might expect, ... and actually a false positive is a much higher cost to the system because then you are referring patients (unnecessarily)"* . Clinician requirements may also dictate trade-offs. For low-risk applications lower accuracy and higher automation maybe desirable for efficiency savings, but presents an unacceptable risk where diagnostic accuracy is critical. This can include solving the problem of where to place the AS in the flow of work where, for example, patients are referred to successively more specialised clinicians.

Our participants emphasised the importance of prospective clinical studies in MD development to provide an evidence-based validation of the accuracy and behaviour of a system in a live working environment. As P4 states *"we don't often approach complex systems as a whole, we dig down into the subcomponents and we understand those in minute detail, but we're a lot (more) hazy as to how, when we put them together, they will act. ... Prospective studies then are very important because we have an opportunity to observe different things working together"*.

This is problematic for SaMD as current regulations restrict approved use of software applications to the hardware systems that they have been tested on, ensuring that the update and upgrade cycles remains within regulatory control and quality managed under the responsibility of a single manufacturer. SaMD that can run on multiple platforms risks obsolescence, uncontrolled change, and the reliance of interdependent vendors ,which as P2 points out is *"quite a tricky ecosystem to keep going"*.

## 3 DISCUSSION

The preliminary findings provide insights into the ongoing challenges for MDR and demonstrate that current models of risk-based governance are intractable to new generations of intelligent technologies. The challenge for governance is to retain the safety integrity assured by current regulation, while recognising the benefits of software-based technologies. This includes the thorny issue of ML software that "learns on the job" which is deemed too risky for the MD market, but has the potential to improve accuracy through continuous training data development. While the current regulatory framework accommodates change processes in hardware-based solutions, just what constitutes a 'substantial' change in terms of software is not fully agreed upon in professional communities.

The MHRA's proposal to include Predetermined Change Control Plans (PCCP) [16] in the MDR framework to track the compliance of a product throughout its lifecycle, would seem to be a step toward rationalising this problem, however a challenge to this approach exists in prospective legislation elsewhere. Article 43 of the EU AIA [3], states that systems that continue to learn after being placed on the market "shall not constitute a significant modification", while Article 3 defines a modification as "a change that affects compliance". This potentially opens up a contradiction for PCCPs. A fundamental premise of ML is the prospect for continuous improvement through feedback of training data, which could be gathered from various sources including data gathered 'on the job' within its context of work. This is theoretically feasible, providing data are validated and fit for purpose – poor quality data or data that is for some reason not a fit for the context of use can cause data drift, or bias, degrading the accuracy and safety integrity of a device. The proposed PCCP ostensibly aims to check for such risky behaviours that manifest while in-service, but which could also potentially render a product as performing outside of the bounds of regulatory compliance. The EU AIA seemingly ignores the introduction of risk from incremental changes to systems that are implemented within the scope of the intended use of a product, but which may in certain cases, retrospectively constitute the legal definition of a modification. This study highlights the regulatory challenges to make intelligent systems achievable in practice, but also the challenge of harmonising approaches that are suited to a specific domain with AI legislation in general.

The studies also highlight social measures of trustworthiness, which is an area in which we aim to track further developments to expand our findings. For example, as we have seen margins of error can differ depending on the placement of AI in the flow of work; the customers' requirements for automation; and the context of use. This raises the question whether trustworthiness may be quantified simply as a measure of accuracy without consideration of these social factors [22]. Smith and Fotheringham [23], for example, argue the risks of Medical AI, including ML that 'learns on the job', "necessitates clinical oversight of the systems' outputs to ensure the AIS's recommendations are safe and relevant to the patient" and discusses AI's potential role in cases of clinical harm or negligence, arguing that the onus of responsibility should be distributed between the Software Developer and the clinician, in terms of duty of care.

## 4 CONCLUSION

In this paper, we explore the prospect of AI in the domain of medical device development and manufacture from two different perspectives, firstly from a research and development team introducing AI driven features to their products, and secondly from the regulator attempting to govern new generation of medical devices within a risk-based framework. Whilst the findings are specific to this domain, there are some general take-aways for the AI community regarding good practices in innovation and responsible product development that is responsive to both legal and ethical obligations. While AI in health care remains a contested topic, the advancements in this area and the legislative response to them provide a 'future echo' for other areas of AI application to heed what might be expect in practical terms when faced with generalised legislation such as the EU's AIA. RI is not something that can be enacted only at an individual level, it will inevitably require the support of research institutions in the university sector and industry.

## ACKNOWLEDGMENTS

This work was supported by the following UK Engineering and Physical Sciences Research Council Grants: UKRI Trustworthy Autonomous Systems Node in Governance and Regulation [grant number EP/V026607/1]; and Horizon: Trusted Data-Driven Products [grant number EP/T022493/1].